# Evidence of valence band perturbations in GaAsN/GaAs(001): A combined variable-angle spectroscopic ellipsometry and modulated photoreflectance investigation

S. Turcotte,[1] S. Larouche,[1] J.-N. Beaudry,[1] L. Martinu,[1] R.A. Masut,[1] P. Desjardins,[1] and R. Leonelli[2]

[1] Département de Génie Physique and Regroupement Québécois sur les Matériaux de Pointe (RQMP), École Polytechnique de Montréal, P.O. Box 6079, Station Centre-ville, Montréal, Québec, Canada H3C 3A7

[2] Département de Physique and Regroupement Québécois sur les Matériaux de Pointe (RQMP), Université de Montréal, P.O. Box 6128, Station Centre-Ville, Montréal, Québec, Canada H3C 3J7

**Abstract**

The contribution of the fundamental gap $E_-$ as well as those of the $E_- + \Delta_{so}$ and $E_+$ transitions to the dielectric function of $GaAs_{1-x}N_x$ alloys near the band edge were determined from variable-angle spectroscopic ellipsometry and modulated photoreflectance spectroscopy analyses. The oscillator strength of the $E_-$ optical transition increases weakly with nitrogen incorporation. The two experimental techniques independently reveal that not only the oscillator strength of the $E_+$ transition but also that of $E_- + \Delta_{so}$ become larger compared to that of the fundamental gap as the N content increases. Since the same conduction band is involved in both the $E_-$ transition and its split-off replica, these results reveal that adding nitrogen in $GaAs_{1-x}N_x$ alloys affects not only the conduction but also the valence bands.

## I. Introduction

The large band gap reduction observed in III–V dilute nitride semiconductors is promising for various optoelectronic devices, such as photovoltaic cells,[1,2] semiconductor lasers,[3,4] photodiodes,[5] and heterojunction bipolar transistors.[6,7] In order to optimize the design of these devices, a better understanding of the optical properties, and, accordingly, of the optical constants (the dielectric function) and of the band structure of these materials is required.

In addition to significantly decreasing the band gap energy, the incorporation of nitrogen in GaAs splits the conduction band into two non-parabolic subbands.[8] A phenomenological model – the band anticrossing (BAC) model – has been successfully used to describe the direct gap transition energies associated with these subbands.[8] The lower transition associated to the first conduction subband is labeled $E_-$, the next higher one $E_+$, and the split-off transition $E_- + \Delta_{so}$. The BAC model is based on the observed apparent repulsion of those bands when the alloy is subjected to hydrostatic pressure.[8] However, it oversimplifies the effects of N in III-V-N alloys by restricting the influence to the $\Gamma$ symmetry point and by assuming that the valence bands remain unaffected.[8] Surprisingly, the validity of this last assumption has not been systematically investigated even though some recent experimental results suggest a possible perturbation of the $GaAs_{1-x}N_x$ valence bands.[9]

Modulation techniques, such as electroreflectance,[10] photoreflectance[11] (PR), and magnetotunneling spectroscopy[12] have been used for direct measurement of the optical transitions, notably $E_+$. Modulated reflectance methods, however, can only provide relative information on the band structure, such as the relative oscillator strength of the optical transitions. Francoeur et al.[10] reported that the $E_+$ transition involves the $L$ critical point (CP) of the band structure. They proposed that breaking the translational symmetry by introducing nitrogen splits the degeneracy of the *L* conduction band and yields new transitions having an $L_{6c}$ symmetry, including the optically active $E_+$. Similar conclusions were also recently reached by Fluegel *et al.*[13]

The strength of the optical transitions in a semiconductor alloy can best be determined experimentally by extracting the dielectric function from spectroscopic ellipsometry measurements. Ellipsometric studies[14-17] of the dielectric function of III-V-N semiconductor alloys have generally been based on a single incidence angle or on the derivation of a pseudo-dielectric function. Recently, our group has shown that variable angle spectroscopic ellipsometry (VASE) has the required sensitivity to detect subtle transitions in $GaAs_{1-x}N_x$, such as the $E_+$ transition, allowing a comparison of the dielectric function with that obtained from first principle calculations.[18] We used a point-by-point fit of the optical constants to confirm that the presence of the $E_+$ related transition, clearly identified in our VASE data, translates into significant contributions to the dielectric function.

In the present article, we report a detailed study of the near-gap optical transitions $E_-$, $E_- + \Delta_{so}$, and $E_+$ from $GaAs_{1-x}N_x$ alloy epilayers grown on GaAs(001). Two complementary optical characterization techniques are used to analyze the effect of nitrogen incorporation on the optical properties of the alloys. The dielectric function is extracted from VASE measurements using a parametric model that is not subject to the noise associated with point-by-point fits. The absorption coefficient at the direct band gap edge, whose value is directly related to the oscillator strength of the lowest energy transition, is also deduced from the imaginary part of the dielectric function. Moreover, modulated photoreflectance measurements are used to independently determine the energy position and the relative oscillator strength of the $E_-$, $E_- + \Delta_{so}$, and $E_+$ transitions. These combined results reveal that not only the oscillator strength of the $E_+$ transition but also that of $E_- + \Delta_{so}$ become larger compared to that of the fundamental gap as the N content increases. Since the same conduction band is involved in both the $E_-$ transition and its split-off replica, these results reveal that adding nitrogen in $GaAs_{1-x}N_x$ alloys affects not only the conduction but also the valence bands.

## II. Experimental details

$GaAs_{1-x}N_x$ samples were grown by organometallic vapor phase epitaxy (OMVPE) in a horizontal low-pressure cold-wall reactor equipped with a fast-switching run-vent manifold with minimized dead volume,[19] using Pd-purified hydrogen as the carrier gas, trimethylgallium (TMGa) as the group-III precursor, and tertiarybutylarsine (TBAs) and 1,1-dimethylhydrazine (DMHy) as group V sources. Growth temperatures of 550 and 575 °C were chosen since they allow controlled nitrogen incorporation over a relatively wide composition range while maintaining sharp optical features. Our samples consist of nominally undoped $GaAs_{1-x}N_x$ layers $0.004 \leq x \leq 0.016$ deposited on a 50-100 nm thick GaAs buffer layer grown on semi-insulating GaAs(001) substrates. The nitrogen composition $x$ was determined from high-resolution X-ray diffraction (HR-XRD) measurements, based on a recent investigation which revealed that Vegard's rule applies in $GaAs_{1-x}N_x$/GaAs(001) samples for up to $x \approx 0.03$.[20] The HR-XRD measurements were carried out in a Philips diffractometer using Cu $K\alpha_1$ radiation ($\lambda = 0.154\,059\,7$ nm) from a four-crystal Ge(220) Bartels monochromator which provides an angular divergence of 12 arc s with a relative wavelength spread $\Delta\lambda/\lambda = 7\times10^{-5}$. $\omega$-$2\theta$ scans (where $\omega$ is the angle of incidence and $\theta$ is the Bragg angle) were obtained with a detector acceptance angle of approximately 2°. Further details concerning the growth procedure and structural characterization can be found in Ref. 20.

PR measurements have been performed at room temperature using a 0.26-m Oriel spectrometer. The monochromatized light of a halogen lamp passing through the spectrometer provided the probe beam. The probe power density at the sample was around 0.2 mW cm$^{-2}$. The wavelength step was set to 5 meV (the resulting bandwidth resolution ranges from 5 meV at 1.35 eV to 9 meV at 1.8 eV). A 25-mW laser emitting at a wavelength of 405 nm was used to modulate the reflectance. Its power density was set between 16 and 140 mW cm$^{-2}$ using neutral density filters in order to minimize Franz-Keldysh effects while providing a sufficiently high signal from the transitions probed.

Ellipsometric measurements were carried out at room temperature using a J.A. Woollam Co. vertical VASE apparatus fitted with a rotating analyzer.[21,22] Since ellipsometry measures the ratio of the parallel to perpendicular amplitude reflection coefficients, no signal modulation is used in this technique, comparatively to PR. A xenon short-arc lamp was used as the light source. A systematic calibration is done at 500 and 1177 nm using a thermally oxidized SiO$_2$/Si reference sample before each measurement run. This procedure ensures the correct determination of the angular position of the input polarizer and of the output analyzer with respect with the plane of incidence and is needed to obtain accurate values of the ellipsometric angles during the spectroscopic experiment. In all cases, a low noise level (typical normalized chi-squared less than 2) was obtained. A relatively large spot size (~1 mm in diameter or more) provided a large signal-to-noise ratio. The measurements cover the available light-source range which extends from 0.76 to 4.40 eV in order to include most optical transitions that are affected by the presence of nitrogen in GaAs$_{1-x}$N$_x$ alloys. The measurement energy step was set to 20 meV or better. For higher energies, the resolution is limited by the 6-nm light bandwidth of the ellipsometric apparatus (corresponding to 3 meV at 0.76 eV and 26 meV at 2.3 eV). Measurements were carried out at four or five angles of incidence near the GaAs Brewster angle (~ 74°); a 0.5° interval between the various angles of incidence was found to result in significant differences in ellipsometric variables $\psi$ and $\Delta$. The samples were cleaned using hot acetone, hot propanol, and deionized water, and dried with nitrogen before optical measurements.

## III. Results

Figure 1 shows typical HR-XRD $\omega$–$2\theta$ scans obtained for the 004 Bragg peak from two samples used in this study (here for $x$ = 0.004 and 0.012). The scans exhibit sharp substrate and layer peaks. Furthermore, finite-thickness interference fringes are clearly visible in both scans, indicating that these alloy layers are of high structural quality with laterally uniform buffer-layer/film interfaces. Simulated HR-XRD scans, based on the fully-dynamical formalism of Takagi[23] and Taupin[24] are shown in Fig. 1 for comparison with experimental data. The simulations were carried out assuming perfectly abrupt and coherent interfaces, and linearly interpolated elastic constants.[25] The measured and simulated curves are in good agreement with respect to the angular position and relative intensities of both the diffraction peaks and the interference fringes. Nitrogen compositions extracted from the analysis of HR-XRD scans assuming Vegard's rule are reported in Table I for all samples. The $GaAs_xN_{1-x}$ layer thicknesses indicated in Table I were determined from HR-XRD (matching the spacing of the finite-thickness interference fringes in experimental and simulated curves) in the case of the thinner samples. For thicker films, thicknesses are obtained by modeling the refractive index dispersion that gives rise to the interference fringes visible in the transparent region of the ellipsometric data.

Fig. 2 presents room temperature PR spectra from the samples listed in Table I. The $E_-$ (and thus the $E_- + \Delta_{so}$) and $E_+$ transitions can be identified since they shift, as expected, to lower and higher energies, respectively, with increasing N content. The absence of Franz-Keldysh oscillations together with the relative changes in reflectance $\Delta R/R$ of $\cong 10^{-4}$ are strong indications that low-field conditions are met in these spectra.[26] The PR lineshapes of each transition can therefore be analyzed using the low-field formula[27,28]

$$\frac{\Delta R}{R} = C\gamma^n \text{Re}\left[e^{i\varphi}(E - E_t + i\gamma)^{-n},\right] \qquad (1)$$

where $C$ is a constant, $\gamma$ is the line broadening, $E_t$ is the transition energy, $\varphi$ is the lineshape phase angle that accounts for the influence of non-uniformity in the electric fields, and $n$ is a parameter that depends on the type of CP. For three-dimensional CPs, $n = 2.5$.[29] The spectral features associated with the $E_-$, $E_- + \Delta_{so}$, and $E_+$ transitions were fitted using Eq. 1 and a least-squares minimization procedure. Experimental and fitted curves are in excellent agreement for all transitions in all samples as depicted in Fig. 3. The parameters obtained for the fits are summarized in Table II.

The energies of the three main transitions extracted from the PR fits are plotted as a function of N content in Fig. 4 and compared with the predictions of the BAC model. The variations of the fundamental band gap and of its split-off replica with N composition are in good agreement with the BAC model predictions. However, while this model predicts that the $E_+$ transition shifts to higher energies in a quadratic way, our results rather indicate a linear blue shift with N incorporation as was also determined from measurements performed at 80 K (Ref. 10) and 10 K.[13] The linear regression intercepts the ordinate axis at $1.75 \pm 0.01$ eV, which lies 1.16 eV below the room temperature GaAs $E_1$ transition energy.[30,31] This value is in agreement with that found by Francoeur *et al.*[10] at a lower temperature and corroborates the *L* nature of the $E_+$ transition.

Experimental VASE data are presented as $\psi$ and $\Delta$ spectra in Fig. 5 for a GaAs substrate and for GaAs$_{1-x}$N$_x$ samples with x = 0.004 to 0.016. The $E_-$ and $E_- + \Delta_{so}$ features are highlighted using gray and black vertical lines, respectively. One should already note that these features are visible for all angles of incidence. While the $E_-$ CP is easily distinguished, the $E_- + \Delta_{so}$ CP is broader and more subtle. As the N content increases, the $E_-$ and $E_- + \Delta_{so}$ transitions red shift and a new feature, indicated with a vertical dashed line, gradually emerges in all $\psi$ and $\Delta$ spectra. This new feature shifts to higher energies with increasing N content while the $E_-$ and $E_- + \Delta_{so}$ CP continue to red shift. No intermediate transition exists between $E_0 + \Delta_{so}$ and $E_1$ in nitrogen-free GaAs and one must therefore consider an additional CP in the analysis of the VASE data from GaAs$_{1-x}$N$_x$ samples. As its energy position closely matches that of $E_+$ observed in PR spectra, we attribute this feature to the $E_+$ transition.

## IV. Discussion

### A. Determination of the dielectric function from VASE data

In order to extract the dielectric function from the VASE data, we developed a bilayer-substrate model reproducing the measured $\psi$ and $\Delta$ spectra acquired at different angles of incidence. The GaAs substrate is assumed to be semi-infinite and comprises the GaAs buffer layer since the optical properties of both layers are essentially identical for the purpose of this analysis. We then consider the dilute nitride layer and an oxide overlayer which we assume to be a typical GaAs native oxide.

The parametric model of Kim and Garland[32] implemented in the WVASE 32 software is used to model the complex dielectric function $\varepsilon = \varepsilon_1 + i\varepsilon_2$ from the VASE spectra. In this framework, each CP is parameterized by its center energy position, its magnitude, and its width, assuming a Gaussian broadening. We begin by fitting the spectra for energies below 2.3 eV for a GaAs substrate covered with a native oxide.[33] The values deduced for the GaAs substrate are then subsequently used for all other steps of the fitting procedure. Then, an initial determination of the $GaAs_{1-x}N_x$ and oxide layer thicknesses is obtained. For thick samples, we modeled the dispersion of the index of refraction at energies below that of the $GaAs_{1-x}N_x$ band gap using the Cauchy dispersion model[34] and fitted the interference fringes to obtain the thickness of the $GaAs_{1-x}N_x$ layers. For thin samples in which these interferences are very broad, we used the thickness obtained from Pendellösung fringes in the HR-XRD spectra as starting values.

The fitting process of the dielectric function of the nitride layer begins with a rough fit of the whole spectra (0.76-4.40 eV) by changing only the $E_1$ and $E_1 + \Delta_1$ transition parameters of the $GaAs_{1-x}N_x$ layer (using the GaAs values as starting points). Since these transitions are clearly affected by N incorporation,[35] this insures that the low energy tail of the $E_1$ related transition will be accurately described. Then, the energy, the amplitude, and the broadening of the three transitions ($E_-$, $E_- + \Delta_{so}$ and $E_+$) occurring below 2.3 eV are allowed to vary. Focusing on that portion of the spectra enables one to concentrate on the relevant transitions and to reduce the number of fitting parameters.[36] For each sample, we fit the spectra at all acquisition angles imposing the same set of parameters. Differences between the measured and calculated $\psi$ and $\Delta$ spectra are minimized using the Levenberg-Marquard algorithm. The quality of the fits is quantified by the normalized chi-squared $\chi_N^2$, whose value stays below 2.0 for every sample. The results of the fitting procedure are presented in Fig. 5. The fits reproduce well the shapes of the different ellipsometric features associated with the three main transitions $E_-$, $E_- + \Delta_{so}$ and $E_+$ for all samples. The resulting fitting parameters are given in Table III. For the sample with $x = 0.004$, the $E_+$ and the $E_- + \Delta_{so}$ transitions are so close in energy that they appear indistinguishable from the raw $\psi$ and $\Delta$ spectra. However, they were resolved by making sure that the fit was also able to reproduce the first three wavelength derivatives of $\psi$ and $\Delta$.

Real ($\varepsilon_1$) and imaginary ($\varepsilon_2$) parts of the dielectric function obtained from modeling the VASE data are presented in Fig. 6 for all GaAs$_{1-x}$N$_x$ samples. The contributions of $E_+$ and $E_- + \Delta_{so}$ to $\varepsilon_2$ increase with nitrogen content. Fig. 6 and Table III also reveal that the contribution of the $E_+$ transition to $\varepsilon_2$ is almost lost for $x = 0.016$. We attribute this behavior to the broadening of the intense $L$-point $E_1$ and $E_1 + \Delta_1$ transitions that already dominate the GaAs $\varepsilon_2$ (up to ~3.5 eV). For instance, focusing on the $E_1$ transition, we measure that the $\varepsilon_2$ maximum is 60 times larger than at the GaAs $E_0$ edge. Upon nitrogen addition to the GaAs lattice, these transitions broaden considerably to become almost undistinguishable, as shown in the inset of Fig. 6(b), in agreement with other ellipsometry studies.[15,35,37] For the sample with $x = 0.016$, the $E_+$ blue shift is the largest and the $E_+$ transition becomes masked by the dominant contributions of the $L$-related transitions. The nitrogen induced breaking of symmetry also leads to a splitting of the first conduction band of the $L$ CP.[10,38] However, the resulting additional transition that would appear near the $E_1$ transition might be difficult to observe due to the weaker effect of nitrogen on the $E_1$ transition than on the $E_-$ and $E_+$ transitions.[39]

The absorption coefficient derived from the dielectric function is plotted as a function of energy in Fig. 7 for all samples. The absolute values are comparable to that of GaAs (~7900 cm$^{-1}$ at threshold)[40] and they slightly increase with nitrogen content over the entire composition range investigated.

**B. Relative oscillator strength**

Examination of Table II reveals that the $E_+$ transition is well distinguished from the split-off transition of the GaAs$_{1-x}$N$_x$ epilayer. As expected, the $E_+$ transition blue-shifts while the $E_-$ and $E_- + \Delta_{so}$ transitions red-shift with nitrogen incorporation. The energy difference between the $E_- + \Delta_{so}$ and $E_-$ transitions, however, slowly decreases with increasing nitrogen, as can also be observed from the data in Ref. 10, at 80 K. In order to further investigate the evolution of these three main transitions, we revisit the PR results, by taking the modulus of Eq. (1)

$$\left| \frac{\Delta R}{R} \right| = \frac{C \gamma^n}{\left[ (E - E_t)^2 + \gamma^2 \right]^{\frac{n}{2}}} . \qquad (2)$$

Eq. (2) describes an $n/2$-order Lorentzian with an approximate width of $2\gamma$. The integrated area is proportional to the PR oscillator strength of the transition.[41] Within a very good approximation, this area is proportional to $C\gamma$. As seen in Table II, it differs widely from sample to sample. To extract quantitative information from this data, it is better to introduce the relative PR oscillator strength $O_{rel}$ of a given transition $t$ with respect to the fundamental band gap transition, which we define as

$$O_{rel} = \frac{C_t \gamma_t}{C_{E_-} \gamma_{E_-}} . \qquad (3)$$

A quantitative evaluation of the oscillator strength of the various transitions can also be extracted from the VASE results. The parametric model described above being an extension of critical-point parabolic band models,[42] the amplitude parameter of the related transition gives the oscillator strength of the transition.[43,44] The relative oscillator strength $S_{rel}$ for a given transition $t$ observed in the ellipsometric measurements is then obtained from the ratio of its amplitude and that of the band edge.

Figure 8(a) presents both relative PR strength $O_{rel}$ (scale to the left) and $S_{rel}$ (right scale) as a function of the nitrogen fraction $x$ for the $E_+$ transition. They both show a rapid rise for $x \sim 0.005$, followed by a lower rate of increase for higher N content. This near-saturation is consistent with *ab initio* calculations of the $\varepsilon_2$ spectrum of $GaAs_{1-x}N_x$.[18] The PR spectra being the result of many convoluted factors involving the reaction of the electronic structure to a modulated electric field, it is not surprising to observe that the actual values of $O_{rel}$ and $S_{rel}$ do not coincide. Nevertheless, the same trend and near-saturation of the oscillator strength is observed for both results.

Figure 8(b) shows both oscillator strengths as a function of $x$ for the $E_- + \Delta_{so}$ transition. The trends shown by the two sets of data are again in very good agreement. Further, the PR values agree well with those reported in Ref. 9. The dielectric function of the GaAs reference sample shown in Fig. 6 is also in good agreement with reported data,[30] the oscillator strength for the split-off transition being one third of that for the fundamental transition. Indeed, using the CP parabolic band model of Adachi, one finds that the oscillator strength is proportional to $\mu^{3/2}$, where $\mu$ is the reduced effective mass, whereas selection rules dictate that the split-off oscillator strength is one half the oscillator strength of the light- and heavy-hole bands.[45,46] Using the combined effective mass of light- and heavy-holes, and the effective mass for the split-off band, one finds that $S_{rel}$ equals 0.34 for GaAs (effective mass values are taken from Ref. 47). This number is very close to the experimental value of 0.32 indicated in Table III.

Both the $S_{rel}$ and $O_{rel}$ values for the $E_- + \Delta_{so}$ transition increase monotonically with $x$, by about a factor of 3 over the N composition investigated. This behavior – which is qualitatively different than that observed in Fig. 8(a) for the $E_+$ transition – is a strong indication that valence bands are also affected by the presence of nitrogen. Indeed, if the valence bands were unaffected by the presence of N, the $E_- + \Delta_{so}$ transition should follow the same trend as the $E_-$ transition since the same conduction band is involved in both ($E_- + \Delta_{so}$ and $E_-$) transitions. The different behaviors depicted in Fig. 8(a) and 8(b) for the $E_+$ and $E_- + \Delta_{so}$ transitions therefore indicate a stronger perturbation of the valence bands with nitrogen than what has previously been envisioned. Our data also suggest that the upper lying valence bands and their split-off replica are affected differently.

A possible origin of the relative increase of the oscillator strength of the $E_- + \Delta_{so}$ transition compared to that of the $E_-$ transition is a decrease of the split-off band curvature. This would imply an increase in the density of states of the split-off band and thus a higher oscillator strength for that transition. Since both the light-hole and split-off band originate from the $m_J = \pm 1/2$ quantum states and couple with the conduction band, a reduction of the split-off band curvature should then also be accompanied by a lowering of the light-hole band curvature. In the 4-band k.p method of Kane for semiconductors with a zincblende structure, for example, a change in $m^*_e$ also leads to a change in the light-hole and heavy-hole effective masses.[48] However, the heavy-hole band being associated with $m_J = \pm 3/2$ quantum states, it does not couple with the first conduction band.[48]

Moreover, the heavy-hole band effective mass is already considerably higher than that of the other two hole bands in the nitrogen free case ($0.53m_0$ for heavy-holes compared to $0.08m_0$ and $0.15m_0$ for light-holes and split-off holes respectively, $m_0$ being the free electron mass).[47] The corresponding heavy-hole density of states is thus dominant. Since the oscillator strength of the $E_-$ transition increases only weakly, as deduced from a comparison of the absorption coefficient spectra of Fig. 7, we conclude that the increase in the oscillator strength of the split-off and $E_+$ transitions is greater than for the $E_-$ transition, where the effects of increasing the light-hole effective mass could be masked by the dominant density of states of the heavy-hole band. Since the heavy-hole band is not involved in the transition between the first conduction band and the split-off band, the modification of the latter band may thus be observed.

## V. Conclusions

Using a combination of PR and VASE measurements, we have investigated the band structure properties of $GaAs_{1-x}N_x$ layers ($0 \leq x \leq 0.016$) on GaAs(001). We find that the oscillator strength of the $E_-$ optical transition increases weakly with nitrogen incorporation. We also observe that the oscillator strength of the $E_+$ and of the $E_- + \Delta_{so}$ transitions become larger compared to that of the fundamental gap as the N content increases. Since the same conduction band is involved in both the $E_-$ transition and its split-off replica, these results reveal that adding nitrogen in $GaAs_{1-x}N_x$ (001) alloys also affects the valence bands, most likely by increasing the effective masses of the light-hole and split-off bands.

## Acknowledgements

The authors acknowledge S. Francoeur for fruitful discussions. This work was supported by the Natural Sciences and Engineering Research Council of Canada (NSERC), the Fonds Québécois de la Recherche sur la Nature et les Technologies (FQRNT), and the Canada Research Chair Program. S. T., J.-N. B. and S. L. are supported by NSERC and FQRNT scholarships.

**TABLES**

Table I. GaAs$_{1-x}$N$_x$ layer parameters.

| x | Thickness from HR-XRD (nm) | Starting value of the thickness from the fit of the interference fringes (nm)[a] | Incident angles covered during VASE measurements (°) |
|---|---|---|---|
| 0 (GaAs) | N / A | --- | 73.6 - 75.6 |
| 0.004 | 253 ± 6 | --- | 73.9 - 75.4 |
| 0.0066 | --- | 2279 ± 5 | 73.8 - 75.3 |
| 0.012 | 243 ± 6 | --- | 73.6 – 75.6 |
| 0.016 | --- | 1206 ± 1 | 73.8 - 75.3 |

[a] Ref. 34

Table II. Fitting parameters ($C$, $\gamma$ and $E_t$) used for the low-field modeling of the main transitions ($E_-$, $E_- + \Delta_{so}$, and $E_+$) visible in PR spectra. The estimated uncertainties in the values for the amplitude and the broadening are ± 2 % and ± 0.8 meV, respectively while they are well within ± 5 meV for the transition energies.

| | $E_-$ ($E_0$ for GaAs) | | | $E_- + \Delta_{so}$ ($E_0 + \Delta_{so}$ for GaAs) | | | $E_+$ | | |
|---|---|---|---|---|---|---|---|---|---|
| $x$ | $C$ | $\gamma$ (meV) | $E_t$ (eV) | $C$ | $\gamma$ (meV) | $E_t$ (eV) | $C$ | $\gamma$ (meV) | $E_t$ (eV) |
| **0.0** | 1.09 | 10.2 | 1.418 | 0.0308 | 25.7 | 1.752 | - | - | - |
| **0.004** | 1.66 | 9.5 | 1.339 | 0.0224 | 19.0 | 1.682 | 0.0189 | 27.1 | 1.784 |
| **0.0066** | 1.30 | 21.7 | 1.295 | 0.0684 | 35.4 | 1.627 | 0.0469 | 39.2 | 1.793 |
| **0.012** | 1.22 | 26.9 | 1.226 | 0.112 | 37.0 | 1.548 | 0.0563 | 42.5 | 1.817 |
| **0.016** | 1.30 | 39.6 | 1.200 | 0.206 | 45.3 | 1.519 | 0.0895 | 51.2 | 1.875 |

Table III. GaAs$_{1-x}$N$_x$ VASE fitting parameters: Amplitude, broadening, and energy position, are given for the three main transitions. The last column gives the relative oscillator strength $S_{rel}$ for the E_ + $\Delta_{so}$ transition. The estimated uncertainties in the values for transition energies are ± 10 meV for the E_ transition and ± 20 meV for the E_ + $\Delta_{so}$ and the E$_+$ transitions. For the amplitudes, the uncertainties are ± 0.005, ± 0.01 and ± 0.004 for the E_, the E_ + $\Delta_{so}$ and the E$_+$ transitions, respectively. The broadening uncertainty is ± 7 meV for the E_ transition. For the subtle E_ + $\Delta_{so}$ and the E$_+$ transitions, it is estimated to be of the order of the broadening value.

| | Amplitude (arb. units) | | | Broadening (meV) | | | Energy position (eV) | | | $S_{rel}$ |
|---|---|---|---|---|---|---|---|---|---|---|
| x | E_ | E_ + $\Delta_{so}$ | E$_+$ | E_ | E_ + $\Delta_{so}$ | E$_+$ | E_ | E_ + $\Delta_{so}$ | E$_+$ | E_ + $\Delta_{so}$ |
| 0 | 0.306 | 0.10 | --- | 20 | 50 | --- | 1.412 | 1.769[a] | --- | 0.32 |
| 0.004 | 0.290 | 0.09 | 0.052 | 41 | 40 | 15 | 1.339 | 1.682 | 1.747 | 0.29 |
| 0.0066 | 0.340 | 0.14[b] | 0.069 | 56 | 33 | 13 | 1.295 | 1.615 | 1.793 | 0.43 |
| 0.012 | 0.314 | 0.18 | 0.077 | 20 | 78 | 27 | 1.226 | 1.548 | 1.856 | 0.56 |
| 0.016 | 0.330 | 0.30 | 0.008[c] | 20 | 77 | 23 | 1.200 | 1.519 | 1.875 | 0.91 |

[a] The uncertainty on this parameter is ± 30 meV.
[b] The uncertainty on this parameter is ± 0.05.
[c] This E$_+$ feature merges with E$_1$ and E$_1$ + $\Delta_1$.

**FIGURE CAPTIONS**

Figure 1. ω-2θ HR-XRD scans through the 004 Bragg peak from $GaAs_{1-x}N_x$/GaAs(001) samples with $x = 0.004$ and 0.012. Fully dynamical simulations of the HR-XRD ω-2θ scans are shown for comparison. Curves are shifted vertically for clarity.

Figure 2. Room temperature PR spectra from a reference GaAs substrate and from $GaAs_{1-x}N_x$/GaAs(001) samples with $x = 0.004$ to 0.016. The $E_-$, $E_-+\Delta_{so}$, and $E_+$ transitions are identified with open, gray, and black circles, respectively. The feature identified by a star for the sample with $x = 0.004$ is related to the band gap edge of the underlying GaAs substrate.

Figure 3. Experimental (solid lines) and fitted (dashed lines) PR spectra from $GaAs_{1-x}N_x$/GaAs(001) samples with $x = 0.004$ to 0.016. The energy scales were selected to cover the (a) $E_-$, (b) $E_- + \Delta_{so}$, and (c) $E_+$ transitions.

Figure 4. Transition energies obtained from the fit of the PR data in Fig. 3 using Eq. (1). Dashed lines correspond to the predictions of the BAC model whereas the solid line is a linear regression of the $E_+$ experimental data.

Figure 5. Experimental (dashed lines) and fitted (solid lines) VASE data for $GaAs_{1-x}N_x$ samples with (a) $x = 0$ (substrate), (b) $x = 0.004$, (c) $x = 0.0066$, (d) $x = 0.012$, and (e) $x = 0.016$. The vertical gray and black lines correspond to the $E_-$ and $E_- + \Delta_{so}$ transitions, respectively. The black dashed line reveals the $E_+$ transition in the dilute nitride samples.

Figure 6. (a) Imaginary and (b) real parts of the dielectric function of $GaAs_{1-x}N_x$ layers. The curves have been vertically shifted for clarity. Lines are guides to the eyes. The inset in (b) shows the imaginary part of the dielectric function for the sample with $x = 0.012$, in a wider energy range. The arrow highlights the peak around which the $L$-related $E_1$ and $E_1 + \Delta_1$ transitions are located.

Figure 7. Absorption coefficient at 300 K at the $E_-$ threshold for $GaAs_{1-x}N_x$ samples deduced from the ellipsometric analysis for A) $x = 0.004$, B) $x = 0.0066$, C) $x = 0.012$, D) $x = 0.016$. The spectrum from the GaAs substrate is also presented for reference.

Figure 8. Relative oscillator strength $O_{rel}$ (solid symbols) and $S_{rel}$ (open symbols) versus nitrogen fraction $x$ for a) the $E_+$ transition and b) the $E_- + \Delta_{so}$ transition.

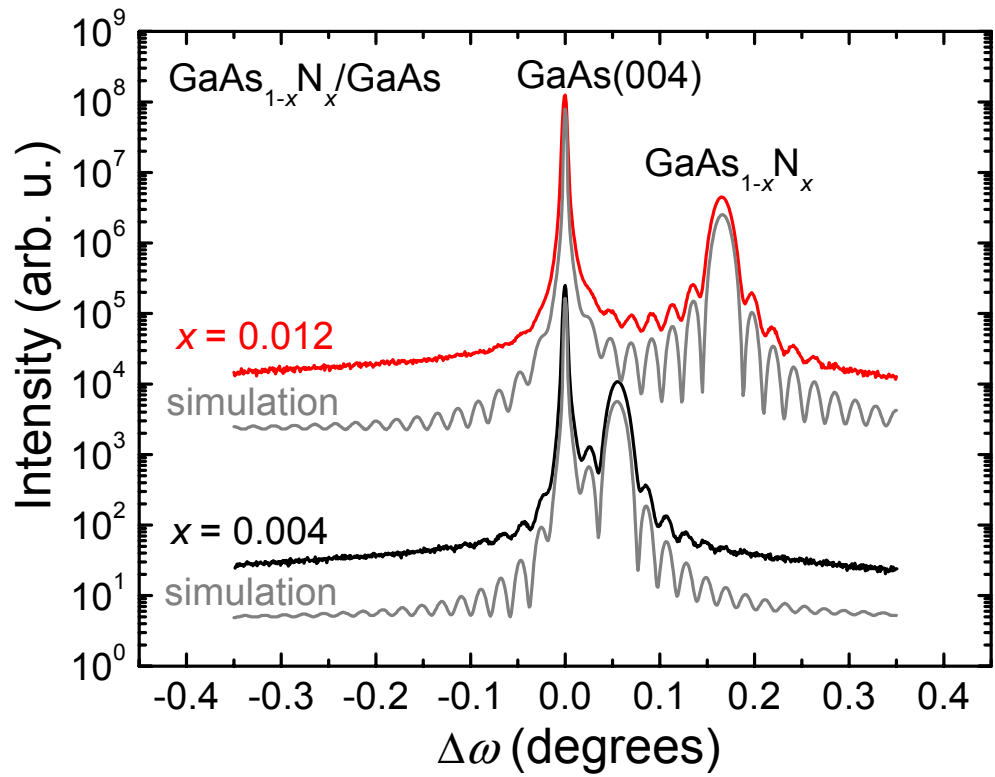

Figure 1
Turcotte et al.

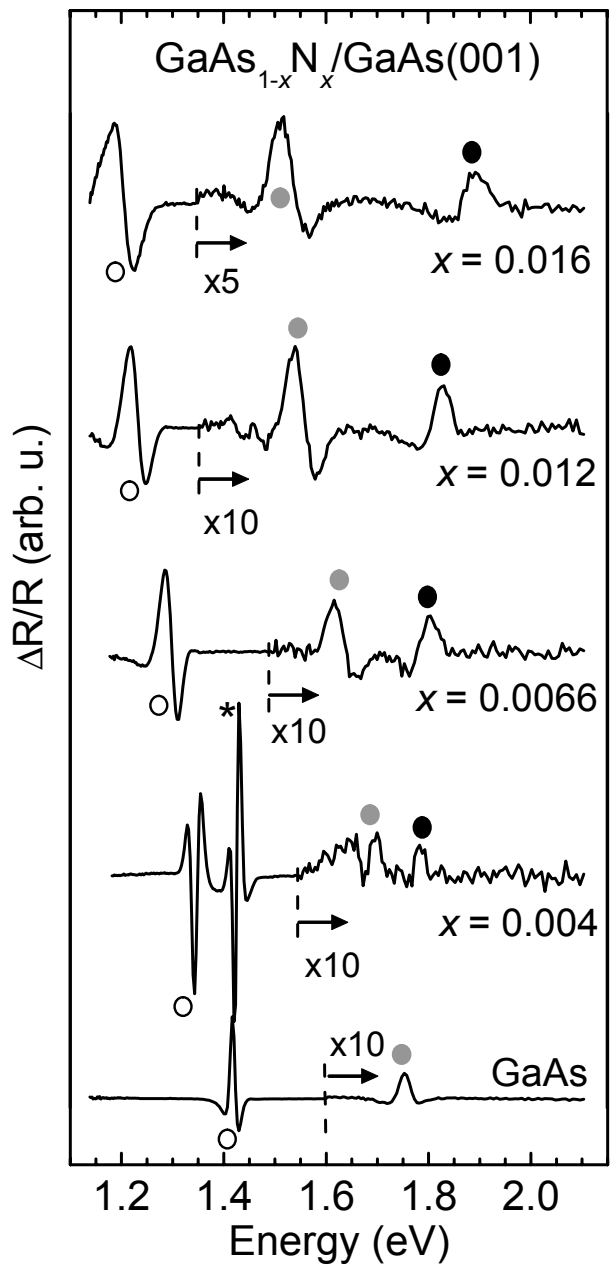

Figure 2
Turcotte et al.

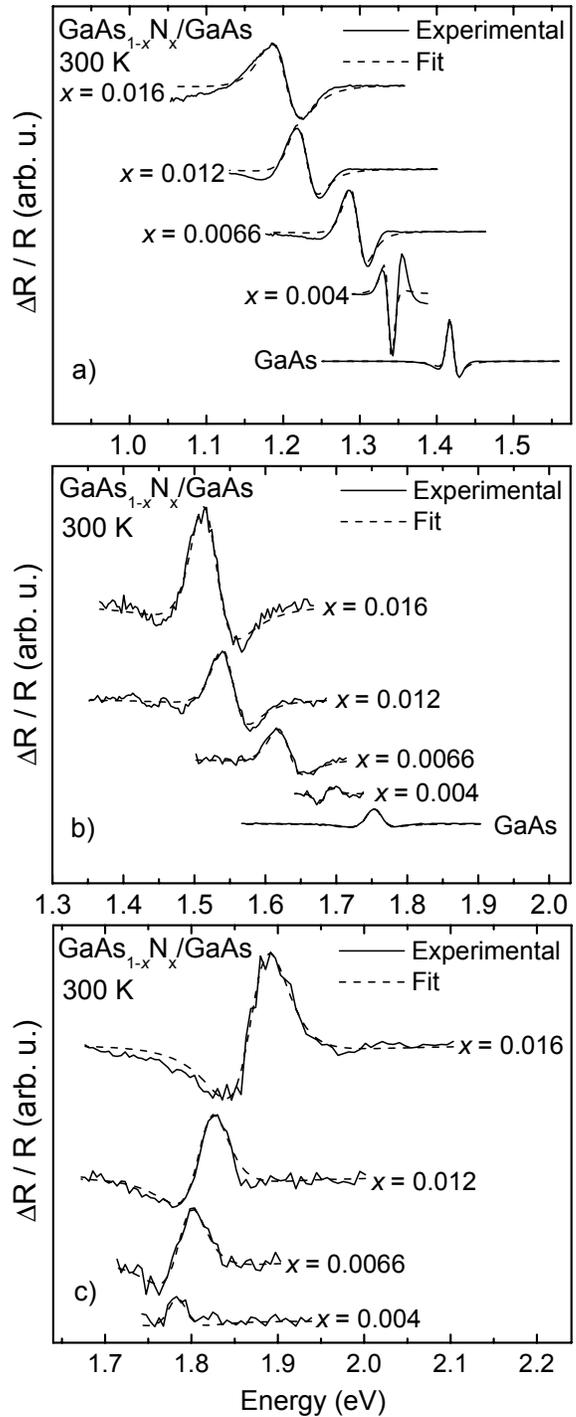

Figure 3
Turcotte et al.

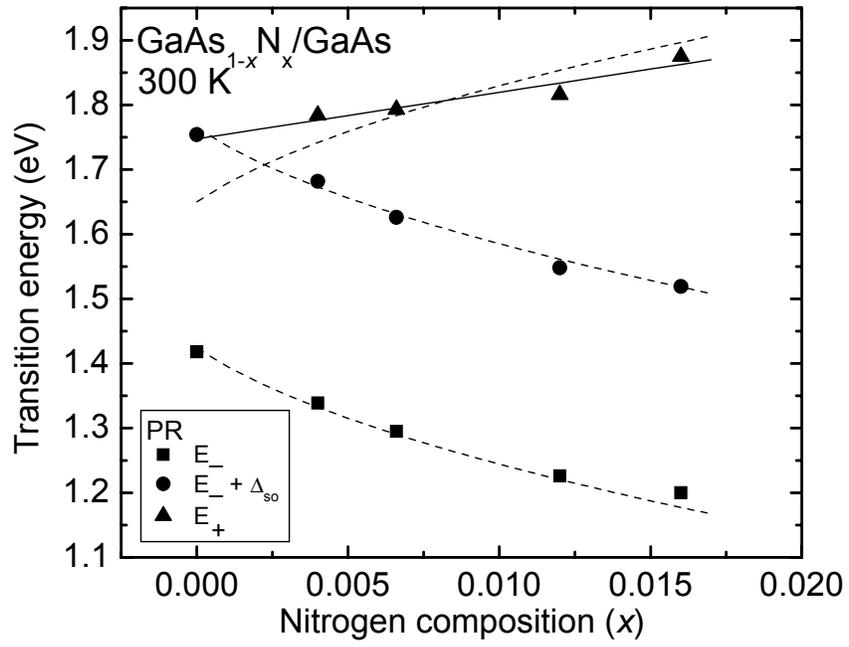

Figure 4
Turcotte et al.

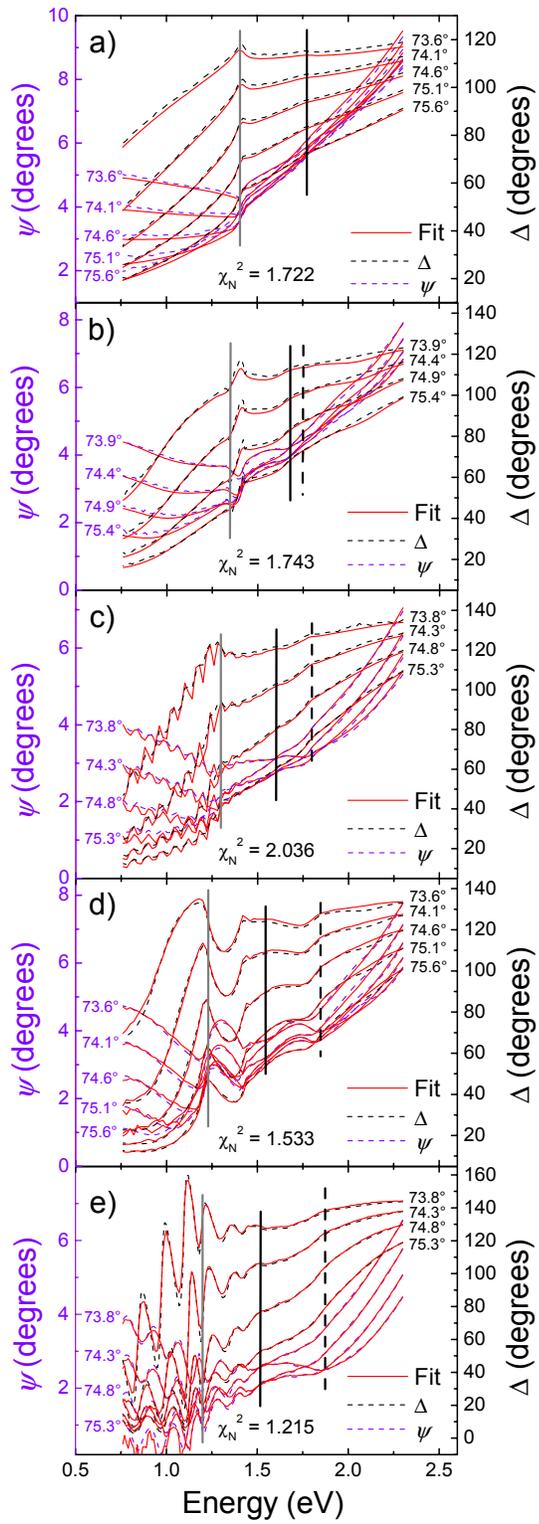

Figure 5
Turcotte et al.

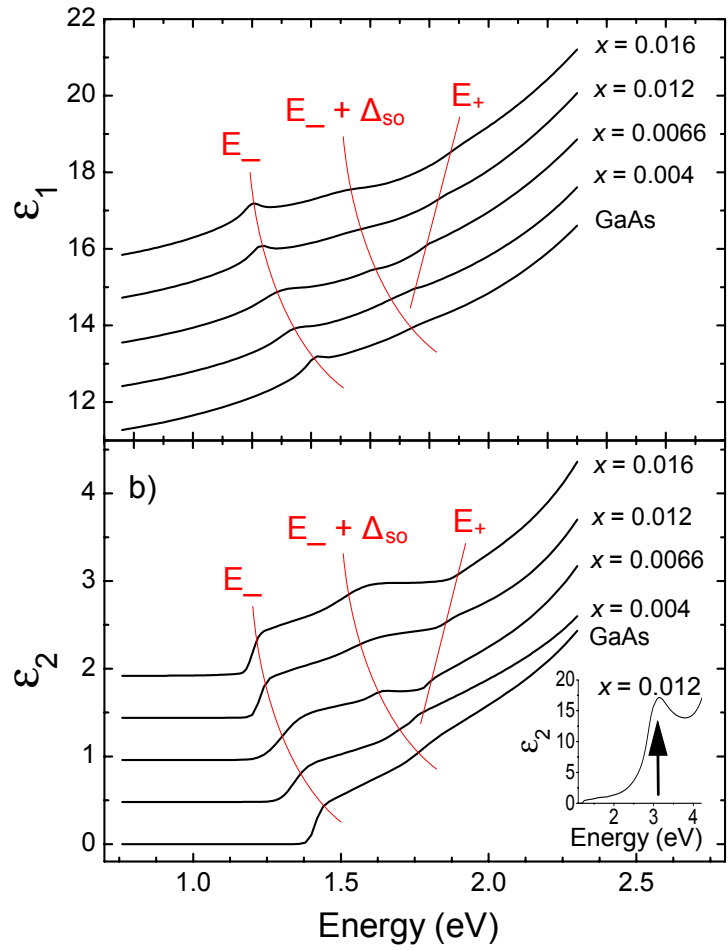

Figure 6
Turcotte et al.

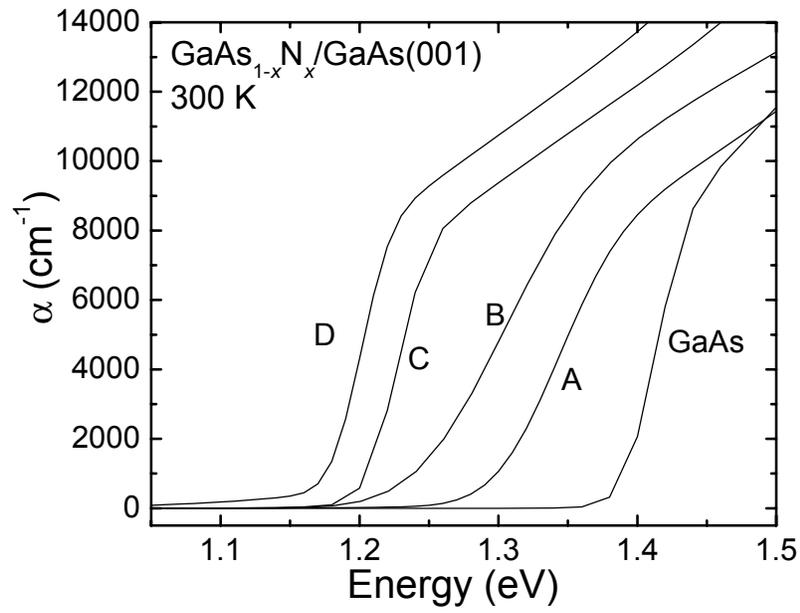

Figure 7
Turcotte et al.

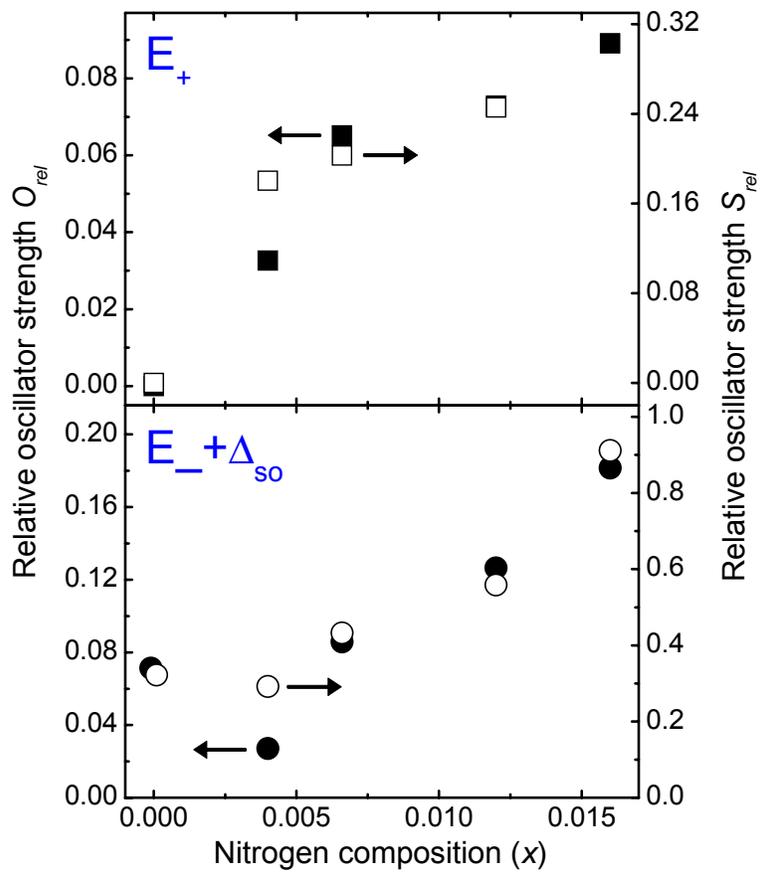

Figure 8
Turcotte et al.